\documentclass[twocolumn,showpacs,preprintnumbers,prl,amsmath,amssymb]{revtex4}
\usepackage{graphicx}

\global\arraycolsep=2pt         

\makeatletter
\def\footnotesize{\@setsize\footnotesize{9.5pt}\xpt\@xpt
\abovedisplayskip 10pt plus2pt minus 5pt
\belowdisplayskip \abovedisplayskip
\abovedisplayshortskip \z@ plus 3pt
\belowdisplayshortskip 6pt plus 2pt minus 2pt
\def\@listi{\topsep 6pt plus 2pt minus 2pt
\parsep 3pt plus 2pt minus 1pt \itemsep \parsep}}
\makeatother

\begin{document}

\title{Highly Charged Ions in a Weakly Coupled Plasma: \\ 
         An Exact Solution }

\author{Lowell S. Brown} \author{David C. Dooling} \author{Dean L. Preston}

\affiliation{
Los Alamos National Laboratory
\\
Los Alamos, New Mexico 87545
\\}

\date{\today}
\begin{abstract}

The ion sphere model introduced long ago by Salpeter is placed in
a rigorous theoretical setting.  The leading corrections to this
model for very highly charged but dilute ions in thermal
equilibrium with a weakly coupled, one-component background
plasma are explicitly computed, and the subleading corrections
shown to be negligibly small.  Such analytic results for very
strong coupling are rarely available, and they can serve as
benchmarks for testing computer models in this limit.

\end{abstract}


\maketitle

Here we shall describe a plasma configuration that is of
considerable interest: very dilute ``impurity'' ions of very high
charge $Z_p e$, $Z_p \gg 1$, in thermal equilibrium with a
classical, one-component ``background'' plasma of charge $ze$ and
number density $n$, at temperature $T = 1/\beta$.  The background
plasma is neutralized in the usual way, and it is weakly coupled.
We use rationalized electrostatic units and measure temperature
in energy units so that the background plasma Debye wave number
appears as $ \kappa^2_D = \beta \, (ze)^2 \, n $.  The weak
coupling of the background plasma is conveyed by $ g \ll 1$,
where $ g = (ze)^2 \, \kappa_D / (4\pi \, T) $.  Although the
background plasma coupling to itself is assumed to be very weak
and the impurity ions are assumed to be so very dilute that their
internal interactions are also very small, we shall require that
the ionic charge $Z_p$ is so great that the coupling between the
impurity ions and the background plasma is large, $g Z_p \gg 1$.
This strongly coupled system is interesting from a theoretical
point of view and our results can be used to check numerical
methods.

This limit can be solved exactly. The solution is given by the
ion sphere result presented by Salpeter \cite{Sal} plus a simple
smaller correction.  This is accomplished by using the effective
plasma field theory methods advocated by Brown and Yaffe
\cite{BY}.  In this field-theory language, the old Salpeter
result corresponds to the tree approximation and our new
correction is the one-loop term.

In usual perturbative expansions, the tree approximation provides
the first, lowest-order term. Here, on the contrary, the tree
approximation provides the leading term for strong coupling, with
the corrections of higher order in the {\it inverse
coupling}. This is the only example of which we are aware in
which the tree approximation yields the strong coupling limit.

Standard methods express the grand canonical partition function
in terms of functional integrals.  Brown and Yaffe \cite{BY} do
this, introduce an auxiliary scalar electrostatic potential, and
(formally) integrate out the charged particle fields to obtain
the effective theory.  The saddle point expansion of this form
for the grand partition function yields a perturbative expansion,
with the tree approximation providing the lowest-order term.
Here, on the contrary, we express the {\it impurity ion number}
in terms of an effective field theory realized by a functional
integral.  This corresponds to a mixed thermal ensemble in which
the very dilute impurity ions are represented by a canonical
ensemble, with the remaining background plasma described by a
grand canonical ensemble. The saddle point of this form of the
functional integral involves a classical field solution driven by
a strong point charge.

The result for the impurity ion number reads
\begin{eqnarray}
N_p &  = &  N_p^{(0)} \, \exp\left\{ {3 \over 10 } \, (3g)^{2/3} \,
 Z_p^{5/3} \right. 
 \nonumber\\
&& \left. + \left({\,9\,\over g}\right)^{1/3} {\cal C} \, Z_p^{2/3}
+ \cdots
- {1\over3} \, g \, Z_p \right\} \,.
\label{num}
\end{eqnarray}
Here $N_p^{(0)} \sim \exp\{ \beta \mu_p \} $ is the the number of
impurity ions defined by the chemical potential $\mu_p$ in the
absence of the background plasma; keeping this chemical potential
fixed, the background plasma alters this number to be $N_p$.  The
final $ - g Z_p / 3$ term in the exponent is the relatively small
one-loop correction.  The added $\cdots$ represent corrections to
the evaluation of the classical action that may or may not be
significant --- if needed, they may be obtained numerically.  The
constant ${\cal C} = 0.8498 \cdots$.

The number correction (\ref{num}) can be used to construct the 
grand canonical partition function ${\cal Z}$ for the combined 
system by integrating the generic relation
$
N = {\partial \over \partial \beta \mu} \, \ln {\cal Z} \,.
$
The equation of
state is then determined from 
$ p V = \ln{\cal Z} $.
To simply bring out the main point, we include here only the leading
terms,
\begin{eqnarray}
p V &\simeq&  \left\{ N - Z_p { (3gZ_p)^{2/3} \over 10} \,
                 \, N_p \right\} \, T \,.
\end{eqnarray}
Although the fraction of impurity ions in the plasma
$ N_p / N$ may be quite small, there may be a significant
pressure modification if $Z_p$ is very large.

The number result (\ref{num}) also directly yields the plasma 
correction to a nuclear fusion rate, since \cite{cite}
\begin{equation}
\Gamma = \Gamma_C \, { N^{(0)}_1 \over N_1} \,
{ N^{(0)}_2 \over N_2 } \,
 { N_{1+2} \over N^{(0)}_{1+2} } \,,
\label{xrated}
\end{equation}
where $\Gamma_C$ is the nuclear reaction rate for a thermal,
Maxwell-Boltzmann distribution of the initial (1,2) particles in
the absence of the background plasma.  We use the notation $1+2$
to denote an effective particle that carries the charge $(Z_1 +
Z_2) e $.  Thus
\begin{eqnarray}
&& \Gamma   =   \Gamma_C  \exp\left\{ {3\over10} \, (3 g)^{2/3} 
\left[ \left( Z_1 \! + \! Z_2 \right)^{5/3} 
\!\! - Z_1^{5/3} \!\! - Z_2^{5/3} \right] \right\}  
\nonumber\\
 && \,\,
 \exp\left\{ \left({\,9\, \over g}\right)^{1/3} {\cal C}
\left[ \left(Z_{1}+Z_{2}\right)^{2/3} - Z_{1}^{2/3}
- Z_{2}^{2/3} \right] \right\} .
\end{eqnarray}
The first line agrees with the calculation of Salpeter
\cite{Sal};the second is new. Again the correction can be large.

We turn now to sketch the basis for these results.  
The effective field theory expresses
\begin{equation}
N_p = { N_p^{(0)} \over {\cal Z}} \, \int [d\chi] \, e^{-S[\chi]} \,,
\label{compact}
\end{equation}
where the effective action is given by 
\begin{eqnarray}
 S[\chi] &  = & 
 \int (d^3{\bf r}) \Bigg\{
{\beta \over 2} \,\Big({\bf \nabla} \chi({\bf r}) \Big)^2
- n \Big[ e^{i \beta z e \, \chi({\bf r})} - 1
\nonumber\\
&& \, \, \,
- i \beta z e \, \chi({\bf r}) \Big] 
- i \beta Z_p e \delta({\bf r}) \, \chi({\bf r}) 
 \Bigg\} \,.
\label{action}
\end{eqnarray}
The normalizing partition function ${\cal Z}$ is the same
functional integral except that the point source $\delta$
function term is removed from the effective action (\ref{action}).
The terms subtracted from the exponential in the action
(\ref{action}) remove an overall number contribution and account
for the effect of the rigid neutralizing background.  As
described in Brown and Yaffe \cite{BY}, one can establish this
result by expanding the exponential in powers of the number
density $n$ and performing the resulting Gaussian functional
integrals to get the usual statistical mechanical form.

The loop expansion is an expansion about the saddle point of the
functional integral.  At this point, the action $S[\chi]$ is
stationary, and the field $\chi$ obeys the classical field
equation. The tree approximation is given by evaluating $S[i
\phi_{\rm cl}({\bf r}) ]$, where $ \phi_{\rm cl}({\bf r})$ obeys
the classical field equation
\begin{equation}
- \nabla^2 \phi_{\rm cl} ({\bf r}) = z e n \left[ 
  e^{-\beta ze \phi_{\rm cl}({\bf r})} - 1 \right]
+ Z e \, \delta({\bf r}) \,.
\label{obey}
\end{equation}
This equation is of the familiar Debye-Huckle form.  We have
placed it in the context of a systematic perturbative expansion
in which the error of omitted terms can be ascertained.  We shall
describe the one-loop correction automatically produced by our
formalism and prove that higher-order corrections may be
neglected.

The one-loop correction to this first tree approximation is
obtained by writing the functional integration variable as
$
\chi({\bf r}) = i \phi_{\rm cl}({\bf r}) + \chi'({\bf r}) \,,
$
and expanding the total action in Eq.~(\ref{compact}) to
quadratic order in the fluctuating field $\chi'$.  Since $i
\phi_{\rm cl}$ obeys the classical field equation, there are no
linear terms in $\chi'$. The leading quadratic terms define a
Gaussian functional integral that produces a Fredholm
determinant.  Hence, to tree plus one-loop order,
\begin{equation}
N_p = N_p^{(0)} \, 
{ {\rm Det}^{1/2} \left[ - \nabla^2 + \kappa^2 \right] \over
{\rm  Det}^{1/2} \left[ - \nabla^2 + \kappa^2 \, 
    e^{-\beta ze \, \phi_{\rm cl}} \right] } \,
\exp\left\{- S[i \phi_{\rm cl}] \right\} \,.
\label{numm}
\end{equation}

To solve the classical field equation (\ref{obey}) in the large
$Z_p$ limit, we note that $\phi_{\rm cl}$ must vanish 
asymptotically, hence Eq.~(\ref{obey}) reduces at large distances 
to the Debye form and thus, for $|{\bf r}|$ large,
\begin{equation}
 \phi_{\rm cl}({\bf r}) \simeq ({\rm const}) \,
   { e^{- \kappa |{\bf r}| } \over |{\bf r}| } \,.
\label{large}
\end{equation}
The coordinate integral of $\nabla^{2} \phi_{\rm cl}$
vanishes by Gauss'
theorem, and from Eq.~(\ref{obey}) we obtain the integral constraint
\begin{equation}
 z n \, \int (d^{3}{\bf r}) \, 
\left[1 -  e^{-\beta e \, \phi_{\rm cl}({\bf r}) } \right]
= Z_p \,.
\label{intc}
\end{equation}

For small $|{\bf r}|$, the point source driving term in the classical
field equation dominates, giving the Coulomb potential solution.
Thus we write
\begin{equation}
\phi_{\rm cl}({\bf r}) = { Z_p e \over 4\pi \, r} \, u(\xi) \,,
\label{udef}
\end{equation}
where
$
\xi = \kappa r \,,
$
and the point driving charge $Z_p e$ is now conveyed in the boundary
condition 
$
u(0) = 1 \,.
$
The large $r$
limit (\ref{large}) requires that 
$ u(\xi) \sim e^{- \, \xi}$ for large $\xi$.

To compute the action corresponding to the classical solution, 
we must first regularize it and remove the 
vacuum self energy of the
impurity ion.
It is not difficult to show that this
gives, on  changing variables to $\xi = \kappa r$,
\begin{eqnarray}
 & & \!\!\!\!\!\!\! 
S_{\rm reg}[i\phi_{\rm cl}]  = - \int_0^\infty d\xi \Bigg\{ 
{ g Z_p^2 \over 2} \, \left( { du(\xi) \over d\xi} \right)^2 
\nonumber  \\ 
&& + { \xi^2 \over g}   \left[ 
\exp\left\{ - { g Z_p \over \xi} \, u(\xi) \right\} - 1 
+ { g Z_p \over \xi} \, u(\xi) \, \right] \Bigg\} \,.
\label{aaction}
\end{eqnarray}
The classical field equation now appears as
\begin{equation}
- g Z_p  \,  { d^2u(\xi) \over d\xi^2} =  \xi \,  \left[ 
\exp\left\{ - { g Z_p \over \xi} \, u(\xi) \right\} - 1 
     \right] \,,
\label{equation}
\end{equation}
according to the variation of Eq.~(\ref{aaction}).

\begin{figure}  
\hspace{-2cm}
\includegraphics[width=7cm,height=4cm]{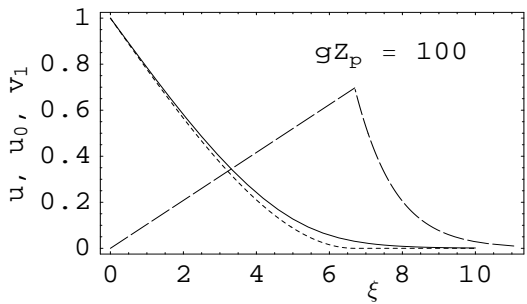}
\hspace{-2.3cm}  
{$\xi$}  
\caption{Numerical solution for $u(\xi)$ (solid line), the 
ion sphere model $u_0(\xi)$ (short-dashed line), and the first 
analytic correction $v_1$ (long-dashed line).
Recall that $u_0 = 0$ for $\xi > \xi_0$; here $\xi_0 = 6.694$.}
\label{uofxi}
\end{figure}

In our large $Z_p$ limit, the short distance
form of Eq.~(\ref{udef}) (multiplied by $\beta ze $) is
large over a wide range of $|{\bf r}|$, 
and 
$ \exp\{ - \beta ze \phi_{\rm cl}({\bf r}) \} $ is quite small there,
  leading to the ``ion sphere model'' introduced long
 ago by Salpeter \cite{Sal}. This model makes the step-function 
approximation 
\begin{equation}
\left[ 1 -
\exp\left\{ - { g Z_p \over \xi} \, u(\xi) \right\} \right] 
\simeq \theta\left( \xi_0 - \xi \right) \,.
\label{step}
\end{equation}
Placing this in the integral constraint (\ref{intc}) determines
the ion sphere radius $\xi_0 = \kappa r_0 $ to be given by
$
 \xi_0^3 = 3 g Z_{p} \,.
$
Approximating Eq.~(\ref{equation}) with the replacement 
Eq.~(\ref{step}) produces a simple differential equation whose 
solution obeying the boundary conditions is
\begin{equation}
u_0(\xi) = \left\{
\begin{array}{ll}
1 - \left( \xi / 2 g Z_p \right) \left[ \xi_0^2 - {1\over3} \xi^2 
              \right] \,, & \mbox{$ \xi < \xi_0 \,, $} 
\\
       0  \,, & \mbox{$ \xi > \xi_0 \,. $}
\end{array}
\right.
\label{answer}
\end{equation}
The nature of this ``ion-sphere'' solution $u_0(\xi)$ together with
the exact solution $u(\xi)$ obtained by the numerical integration of
Eq.~(\ref{equation}), as well as the first correction described below, 
are displayed in Fig.~\ref{uofxi}.  

We have appended the subscript $0$ to indicate that this is the
solution for the ion sphere model.  Placing this solution in the
new version (\ref{aaction}) of the action gives
\begin{equation}
- S_0\left[ i \phi_{\rm cl} \right] 
= {3 Z_p \over10} \,  (3 g Z_p)^{2/3}  - Z_p \,.
\label{sphact}
\end{equation}

To find the leading correction, we
first write the full solution $u(\xi)$ as
$u(\xi) = u_0(\xi) + ( \xi_0 / Z_p g ) \, v(\xi)$,
where $u_0(\xi)$ is the solution (\ref{answer}) to the ion sphere
model.  The exact differential equation
(\ref{equation}) now reads
\begin{eqnarray}
 { d^2 v(\xi) \over d\xi^2 } &  = &  {\xi \over \xi_0} 
\left[ \theta\left( \xi - \xi_0 \right) -
 e^{- \frac{g Z_p u_0(\xi)}{ \xi} }  \exp\left\{ 
  -  { \xi_0 \over \xi}  v(\xi) \right\} 
      \right] .
\nonumber\\
&&
\label{neweq}
\end{eqnarray}
Since $u_0(0) = 1$ and since the solution must vanish at infinity,
the proper solution to Eq.~(\ref{neweq}) obeys
$
v(0) = 0 \,, 
$
and
$
 v(\xi) \to 0 
$
for 
$
\xi \to \infty
$.
Some algebra yields
\begin{eqnarray}
S_{\rm reg}[i\phi_{\rm cl}] &=& S_0[i\phi_{\rm cl}] - 
{ \xi_0 \over g } \, \int_{\xi_0}^\infty d\xi \, \xi \, v(\xi) 
\nonumber \\
&& \quad    - { \xi_0^2 \over 2g} \, \int_0^\infty d\xi \,
        \left( { dv(\xi) \over d\xi } \right)^2 \,.
\label{aaaction}
\end{eqnarray}

Thus far we have made no approximations.  To obtain the leading
correction to the ion sphere result, we note that the factor
$ \exp\left\{ - (g Z_p / \xi) \, u_0(\xi) \right\} $
is very small for $ \xi < \xi_0 $, and so it may be evaluated by
expanding $u_0(\xi)$ about $\xi = \xi_0$. Using  
Eq.~(\ref{answer}),  the leading terms yield
\begin{displaymath} 
\exp\left\{ - { g Z_p \over \xi} \, u_0(\xi) \right\} \simeq
\exp\left\{ - {1\over2} \, \left( \xi_0 - \xi \right)^2 \,
    \theta\left( \xi_0 - \xi \right)  \right\} \,.
\end{displaymath}
This approximation is valid for all $\xi$ because when $\xi$ is
smaller than $\xi_0$ and our expansion
breaks down, the argument in the exponent is so large that
the exponential function essentially vanishes. Since we
consider the limit in which $\xi_0$ is taken to be very large and
the Gaussian contribution is very narrow on the scale set by
$\xi_0$, we may approximate    
\begin{equation}
\exp\left\{ - {g Z_p \over \xi} \, u_0(\xi) \right\} \simeq
\sqrt{ \pi \over 2} \, \delta\left( \xi - \xi_0 \right) +
   \theta\left(\xi - \xi_0 \right) \,.
\label{leader}
\end{equation}
Here the delta function accounts for the little piece
of area that the Gaussian provides near the ion sphere radius.
One may verify that, with
this approximation
inserted into Eq.~(\ref{neweq}), the leading solution  
is given for
$\xi < \xi_0$ by   $v_1(\xi) = c_1 \, \xi \,$ ,
where $c_1$ is a constant that is yet to be determined, while
\begin{equation}
\xi > \xi_0 \,: \qquad\qquad 
{ d^2 v_1(\xi) \over d \xi^2} = 1 - e^{- v_1(\xi) } \,.
\label{ddiffeq}
\end{equation}
This differential equation is akin to a one-dimensional
equation of motion of a particle with $\xi$ playing the role of
time, and $v_{1}(\xi)$ playing the role of position.
Thus there is the usual ``energy constant of motion''.
The integration constant is fixed by requiring that $v_1(\xi)$
vanishes at infinity. Then  
choosing the proper root to ensure that 
asymptotically $v_1(\xi)$ decreases when $\xi$ increases gives
\begin{equation}
 {  d v_1(\xi) \over d\xi } = - \sqrt{ 2 \, \left[ 
 e^{ - \, v_1(\xi) } +v_1(\xi) - 1 \right] } \,.
\label{slope}
\end{equation}

The different functional forms for $v_1(\xi)$ in the two regions 
$\xi < \xi_0$ and $\xi > \xi_0$ are joined by the continuity
constraint 
$
c_1 \, \xi_0 = v_1(\xi_0) \,,
$
and a slope jump to produce the $\delta$ function introduced 
by Eq.~(\ref{leader}). This requires 
\begin{equation}
 \sqrt{ 2 \, \left[  e^{ - \, v_1(\xi_0) } +v_1(\xi_0) - 1 \right] } 
= \sqrt{\pi \over 2} \, e^{- v_1(\xi_0) } - {v_1(\xi_0) \over \xi_0} 
     \,.
\end{equation}
For $\xi_0 \gg 1$, the second term on the right-hand
may be neglected, giving
$
v_1(\xi_0) = 0.6967 \cdots  \,.
$

We now evaluate the leading correction in the
action (\ref{aaaction}). 
In computing the leading term we can set $\xi= \xi_0$ 
in the integral that is linear  in $v_1(\xi)$. The
leading correction  is given by
$
S_{\rm reg}[i\phi_{\rm cl}] \simeq S_0[i\phi_{\rm cl}] + S_1 \,,
$
in which
$
S_1 = - (\xi_0^2 / g ) \, {\cal C} \,,
$
where
\begin{equation}
{\cal C} = \int_{\xi_0}^\infty d\xi \left\{ v_1(\xi) + {1\over2} \,
  \left( { d v_1(\xi) \over d\xi } \right)^2 \right\} \,.
\label{calc}
\end{equation}
Here we have omitted a part that is of
the negligible relative order 
$1 / \xi_0$.  We change variables
from $\xi$ to $v_1$
and use the result (\ref{slope}) for the derivative
to get simple numerical integrals yielding
$
{\cal C} = 0.8498 \cdots \,.
$

\begin{figure}
\hspace{-2cm}
\includegraphics[width=7cm,height=4cm]{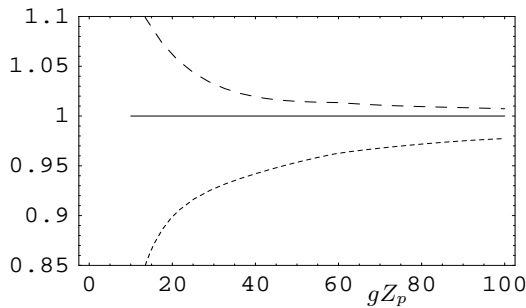}
\hspace{-2.4cm} {$gZ_{p}$} 
\caption{Ratios of the modified action $S[i\phi_{\rm cl}] - Z_{p}$ for 
the ion sphere model result (\ref{sphact}) [short-dashed line]
and the corrected ion sphere model (\ref{next}) [long-dashed line] 
to the action (\ref{aaction}) of the exact numerical solution 
as functions of $gZ_{p}$.}
\label{ratio}
\end{figure}

In summary, we now find that
\begin{equation}
 - [S_0 + S_1] + Z_p  = 
{ 3 Z_p \over 10 } \, \left( 3gZ_p \right)^{2/3} \,
    \left\{ 1 + { 10 \, {\cal C} \over \left( 3 g Z_p ) \right) }
   \right\} \,. 
\label{next}
\end{equation} 
The leading correction to the ion sphere model is
of relative order $ 1 / ( gZ_p) $. Fig.~\ref{ratio} displays the exact 
numerical evaluation of the action compared with the ion sphere
approximation and the corrected 
ion sphere model.

The one-loop correction for the background plasma with no ``impurity''
ions present is given by \cite{foot}
\begin{equation}
{\rm Det}^{-1/2} \left[ - \nabla^2 + \kappa^2 \right] = 
\exp\left\{ \int (d^3{\bf r}) \, { \kappa^3 \over 12 \pi} \right\} \,.
\label{akin}
\end{equation}
We assume that the charge $Z_p$ of the ``impurity'' ions is so
large that not only $Z_p \gg 1$, but also $ g Z_p \gg 1$ as well, even
though we require that $ g \ll 1$.  Then $ \kappa r_0 \gg 1$, and the 
ion sphere radius $r_0$ is large in comparison to the characteristic
distance scale for  spatial variation in the background plasma, the
Debye length $ \kappa^{-1}$.  In  this case, the term
$
\kappa^2 \, \exp\left\{ - \beta ze \phi({\bf r}) \right\} 
$
in the one-loop determinant 
$
{\rm Det}^{-1/2} \left[ - \nabla^2 + \kappa^2 \,
      e^{- \beta ze \phi_{\rm cl} } \right]
$
in Eq.~(\ref{numm}) 
can be treated as being very slowly varying --- essentially a constant
--- except when it appears in a final volume integral akin to
that in Eq.~(\ref{akin}). Therefore, for very strong coupling,
\begin{eqnarray}
&&  \!\!\!\!\!\!\!\!
 { {\rm Det}^{1/2} \left[ - \nabla^2 + \kappa^2 \right] \over 
{\rm Det}^{1/2} \left[ - \nabla^2 + \kappa^2 \,
      e^{- \beta ze \phi_{\rm cl} } \right]  }
\nonumber \\
&& =
\exp\left\{-  { \kappa^3 \over 12 \pi} \, \int (d^3{\bf r}) \,
\left[ 1 - \exp\left\{ - \beta ze \phi({\bf r}) \right\} \right]
\right\} 
\nonumber\\
&& =
\exp\left\{-  { \kappa^3 \over 12 \pi} \, {4\pi \over 3} \, 
r_0^3 \right\} 
 = \exp\left\{ - {1\over3} \, g Z_p \right\} \,,
\label{oneloop}
\end{eqnarray}
where in  the  second equality we have used the ion sphere  model that
gives the leading term for large $Z_p$. 

This result is physically obvious.  The impurity ion
carves out a hole of radius $r_0$ in the original, background
plasma.
The original, background plasma is unchanged outside
this hole. The corrections that smooth out
the sharp boundaries in this picture only produce higher-order
terms.  The original, background plasma had a vanishing
electrostatic potential everywhere, and the potential in the ion
sphere picture vanishes outside the sphere of radius $r_0$.
The grand potential of the background plasma is now reduced
by the amount originally contained within the sphere of
radius $r_0$, and this is exactly what is stated to one-loop
order in Eq.(\ref{oneloop}). 

This argument carries on to the higher loop terms as well. 
As shown in  detail in the paper of Brown and Yaffe \cite{BY},
$n$-loop terms in the expansion of the background plasma partition
function with no impurities present involve a factor of  
$ \kappa^2 \, \kappa^n $ which combines with other
 factors to give dimensionless terms of the form
$
g^{n-1} \, \int (d^3{\bf r}) \, \kappa^3 \,.
$
With the very high $Z_p$ impurity ions present, each factor of
$\kappa$ is accompanied by 
$\exp\{ - (1/2) \beta z e \, \phi_{\rm cl}({\bf r}) \} $
whose spatial variation can be neglected except in the final
volume integral. Thus, 
 an $n$-loop term is of order
\begin{eqnarray}
&& g^{n-1} \kappa^3 \, \int (d^3{\bf r}) \,
\left[ 1 - \exp\left\{ - {n + 2 \over 2} \, \beta z e 
          \phi_{\rm cl}({\bf r}) \right\} \right]
\nonumber \\
&& \qquad\qquad\qquad
\sim g^{n-1} \, \kappa^3 r_0^3 \sim g^n \, Z_p \,.
\end{eqnarray}
Since we assume that $g$ is sufficiently small 
that $g^2 \, Z_p \ll 1 $ (even though $ g Z_p \gg 1$), 
all the higher loop terms may be neglected \cite{fooot}.

We have now established the results quoted above. 

We thank Hugh E. DeWitt and Lawrence G. Yaffe for providing 
constructive comments on preliminary versions of this work.


\begin{thebibliography}{99}
\bibitem{foot}
     See Eq.~(2.79) of Brown and Yaffe \cite{BY} and the 
     discussion leading to that result.
\bibitem{fooot}
      We have glossed over powers of $\ln g$ and quantum corrections
      that appear in higher orders. They vanish in our strong
      coupling limit.
\end{thebibliography}
\end{document}